\documentclass[pdflatex,sn-mathphys-num]{sn-jnl}% Math and 
\usepackage{graphicx}%
\usepackage{multirow}%
\usepackage{amsmath,amssymb,amsfonts}%
\usepackage{amsthm}%
\usepackage{mathrsfs}%
\usepackage[title]{appendix}%
\usepackage{xcolor}%
\usepackage{textcomp}%
\usepackage{manyfoot}%
\usepackage{booktabs}%
\usepackage{algorithm}%
\usepackage{algorithmicx}%
\usepackage{algpseudocode}%
\usepackage{listings}%
\theoremstyle{thmstyleone}%
\theoremstyle{thmstyletwo}%

\theoremstyle{thmstylethree}%

\begin{document}

\title{A Comprehensive Review of Human Error in Risk-Informed Decision Making: Integrating Human Reliability Assessment, Artificial Intelligence, and Human Performance Models}

\author[1]{\fnm{} \sur{Xingyu Xiao}}\email{xxy23@mails.tsinghua.edu.cn}

\author[2]{\fnm{} \sur{Hongxu Zhu}}\email{zhuhx3@foxmail.com}

\author[1]{\fnm{} \sur{Jingang Liang}}\email{jingang@tsinghua.edu.cn;Tel: +86-158-0162-5345}

\author[1]{\fnm{} \sur{Jiejuan Tong}}\email{tongjj@tsinghua.edu.cn}

\author[1]{\fnm{} \sur{Haitao Wang}}\email{wanght@tsinghua.edu.cn}

\affil[1]{\orgdiv{Institute of Nuclear and New Energy Technology}, \orgname{Tsinghua University}, \orgaddress{\city{Beijing}, \postcode{100084}, \state{Beijing}, \country{China}}}

\affil[2]{\orgdiv{School of Journalism and Communication}, \orgname{Tsinghua University}, \orgaddress{\city{Beijing}, \postcode{100084}, \state{Beijing}, \country{China}}}

\abstract{

Human error remains a dominant risk driver in safety‑critical sectors such as nuclear power, aviation, and healthcare, where seemingly minor mistakes can cascade into catastrophic outcomes. Although decades of research have produced a rich repertoire of mitigation techniques, persistent limitations: scarce high‑quality data, algorithmic opacity, and residual reliance on expert judgment, continue to constrain progress. This review synthesizes recent advances at the intersection of risk‑informed decision making, human reliability assessment (HRA), artificial intelligence (AI), and cognitive science to clarify how their convergence can curb human‑error risk. We first categorize the principal forms of human error observed in complex sociotechnical environments and outline their quantitative impact on system reliability. Next, we examine risk‑informed frameworks that embed HRA within probabilistic and data‑driven methodologies, highlighting successes and gaps. We then survey cognitive and human‑performance models, detailing how mechanistic accounts of perception, memory, and decision‑making enrich error prediction and complement HRA metrics. Building on these foundations, we critically assess AI‑enabled techniques for real‑time error detection, operator‑state estimation, and AI‑augmented HRA workflows. Across these strands, a recurring insight emerges: integrating cognitive models with AI‑based analytics inside risk‑informed HRA pipelines markedly enhances predictive fidelity, yet doing so demands richer datasets, transparent algorithms, and rigorous validation. Finally, we identify promising research directions, coupling resilience engineering concepts with grounded theory, operationalizing the iceberg model of incident causation, and establishing cross‑domain data consortia, to foster a multidisciplinary paradigm that elevates human reliability in high‑stakes systems.

}


\maketitle

\section{Introduction}
\label{Introduction}
In safety-critical sectors like nuclear power, aviation, and healthcare, human error remains a predominant risk driver, where seemingly minor mistakes can escalate into catastrophic outcomes. Despite decades of research yielding a rich repertoire of mitigation techniques, persistent limitations, such as scarce high-quality data, algorithmic opacity, and an ongoing reliance on expert judgment, continue to impede progress in effectively curbing this risk. Understanding and mitigating human error is therefore paramount for enhancing system safety, reducing operational risks, and preventing disastrous events in these high-reliability organizations (HROs).

To address this challenge, risk-informed decision-making (RIDM) has emerged as a crucial approach for evaluating and managing human error. These methods integrate rigorous risk analysis with system performance assessments, enabling organizations to prioritize interventions based on the likelihood and potential consequences of human error. Rather than solely focusing on error elimination, risk-informed approaches aim to identify which errors pose the greatest threats to system safety, allowing for more targeted and effective interventions \cite{amendola2002recent}. This shift towards a probabilistic understanding of human error enhances the ability to design more resilient systems capable of withstanding and recovering from human-related failures.

Human reliability assessment (HRA) is a core component within risk-informed frameworks. Techniques like THERP (Technique for Human Error Rate Prediction) \cite{swain1974human} and CREAM (Cognitive Reliability and Error Analysis Method) \cite{hollnagel1998cognitive} are employed to quantify the likelihood of human error in specific tasks and operational environments. By rigorously assessing how human behavior interacts with complex system dynamics, HRA provides invaluable insights into human error probabilities, thereby pinpointing areas where human performance is most vulnerable to failure. These assessments directly inform RIDM frameworks, ensuring that human factors are thoroughly considered in the design of safety protocols and operational procedures.

Concurrently, artificial intelligence (AI) and human performance models (HPMs) have become critical tools for bolstering safety and reliability in modern complex systems. AI offers an unparalleled capacity to analyze vast amounts of operational data in real time. By processing information from sensor networks, wearables, and human-machine interfaces, AI systems can monitor operator performance and detect subtle indicators of cognitive overload, fatigue, or stress, conditions known to significantly increase error likelihood. Furthermore, AI supports adaptive decision support systems \cite{xiao2024emergency}, providing real-time, context-aware guidance to human operators based on their current cognitive state and the evolving operational environment. AI's continuous learning capabilities from operational data also enhance its predictive accuracy over time, making it an invaluable asset in dynamic, high-risk settings. In parallel, HPMs provide mechanistic accounts of the cognitive, psychological, and behavioral factors that influence human error. These models simulate how operators respond to stressors, workload, and complex decision-making tasks, enabling organizations to design systems and workflows that are inherently aligned with human capabilities and limitations. The integration of HPMs with AI-driven monitoring thus promises to develop more resilient human-machine systems that significantly minimize the impact of human error and elevate overall safety.

\begin{figure}[h]
\centering
\includegraphics[width=1.0 \textwidth]{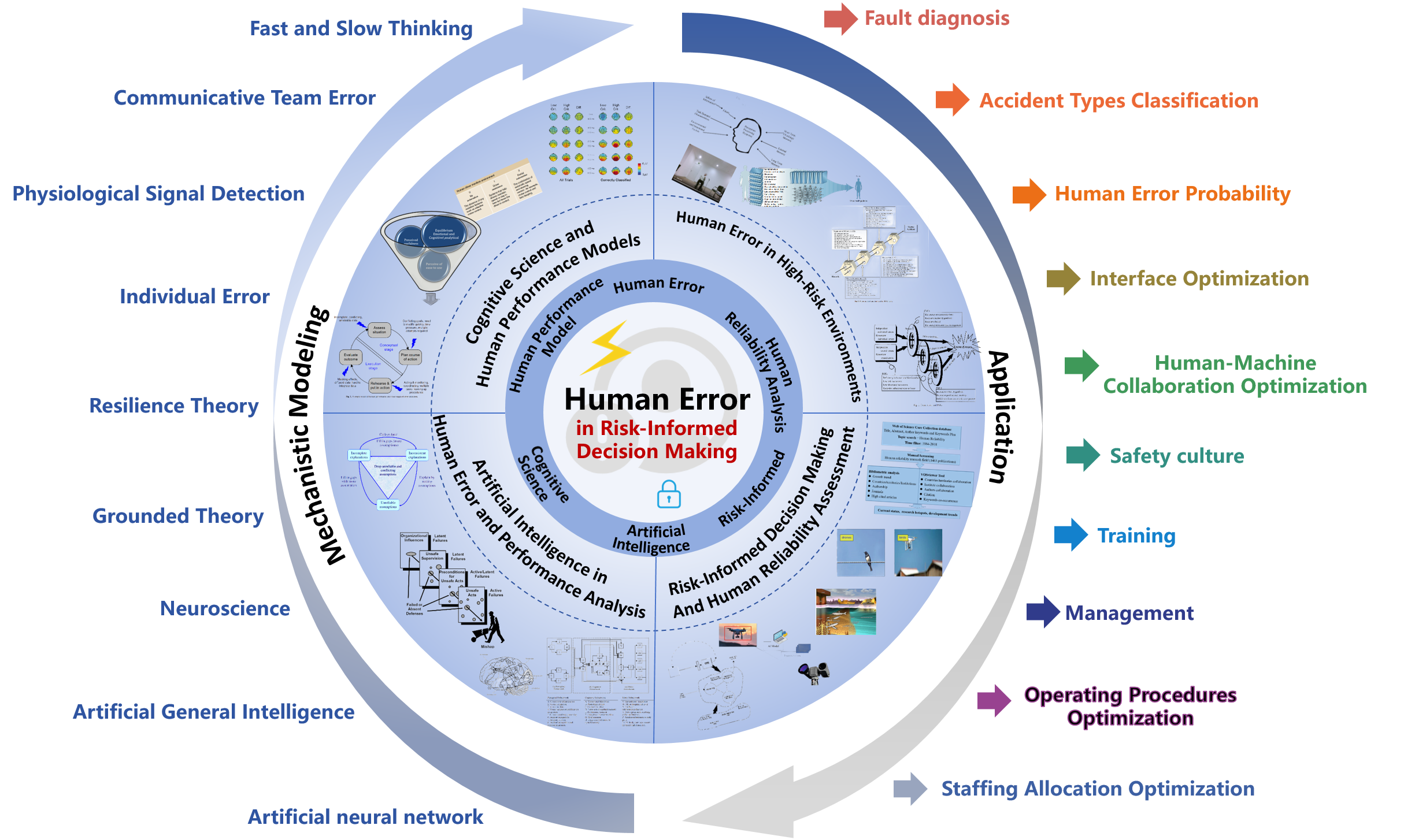}
\caption{Conceptual Framework of Human Error in Risk-Informed Decision Making: Integrating HRA, AI, and Human Performance Models}\label{literature}
\end{figure}

This paper provides a comprehensive review synthesizing recent advances at the intersection of risk-informed decision-making, human reliability assessment, artificial intelligence, and cognitive science. Our aim is to clarify how their convergence can curb human-error risk. The overall framework of this review is illustrated in Figure \ref{literature}. The inner circle highlights six key concepts that form the foundation of our discussion: human error, human reliability, risk-informed approaches, artificial intelligence, cognitive science, and human performance models. The middle layer outlines the paper's main sections: Section 2 categorizes the principal forms of human error observed in complex sociotechnical environments and outlines their quantitative impact on system reliability. Section 3 examines risk-informed frameworks that embed HRA within probabilistic and data-driven methodologies, highlighting successes and gaps. Section 4 surveys cognitive and human performance models, detailing how mechanistic accounts of perception, memory, and decision-making enrich error prediction and complement HRA metrics. Building on these foundations, Section 5 critically assesses AI-enabled techniques for real-time error detection, operator-state estimation, and AI-augmented HRA workflows. Section 6 addresses the current challenges in human error analysis, while the final section concludes with promising research directions. The left side of Figure \ref{literature} presents mechanistic models for quantifying human reliability, such as fast and slow thinking and communicative team errors, while the right side focuses on practical applications like fault diagnosis to enhance system safety and prevent accidents.

\section{Human Error in High-Risk Environments}
\label{Human Error in High-Risk Environments}

Understanding the nature of human error, its consequences, and effective strategies for mitigating it is essential for enhancing overall safety and reliability. The following subsections provide a comprehensive analysis of the definition and classification of human error, explore the profound impact these errors have on system safety, and review the most effective mitigation strategies to minimize their occurrence and consequences.

A statistical analysis of Web of Science documents (Table \ref{tab1_human_error}) shows that terms such as 'root cause,' 'safety engineering,' and 'failure modes and effects analysis' occur infrequently, while keywords related to human factors and ergonomics dominate, highlighting their central role in mitigating human error. The prominence of 'smart' reflects growing integration of AI, automation, and intelligent systems in error prediction and prevention. Frequent mentions of 'benchmarking' and 'performance indicator' underscore the shift toward standardized evaluation methods for assessing human performance and system reliability. Although less frequent, terms like 'human error' and 'human reliability' remain foundational, indicating continued core interest. Overall, research in this domain increasingly combines human-centered design, intelligent technologies, and standardized assessment, reflecting a multidisciplinary and technology-driven evolution.

\begin{table}[h]
\caption{Frequency Analysis of Key Topics Related to 'Human Error' from Web of Science Literature}\label{tab1_human_error}%
\begin{tabular}{p{7cm} p{5cm}}
\toprule
Key Word & Times  \\
\midrule
human factors and ergonomics   &30,156  \\
smart  &22,334       \\
benchmarking  &  15,292   \\
workload  & 14,513     \\
performance indicator    & 9,436    \\
systems design   & 9,195     \\
performance improvement   & 4,328     \\
human error & 2,852 \\
human-centered design & 1,960 \\
human reliability & 1,351 \\
\botrule
\end{tabular}
\end{table}

\begin{figure}[h]
\centering
\includegraphics[width=0.9\textwidth]{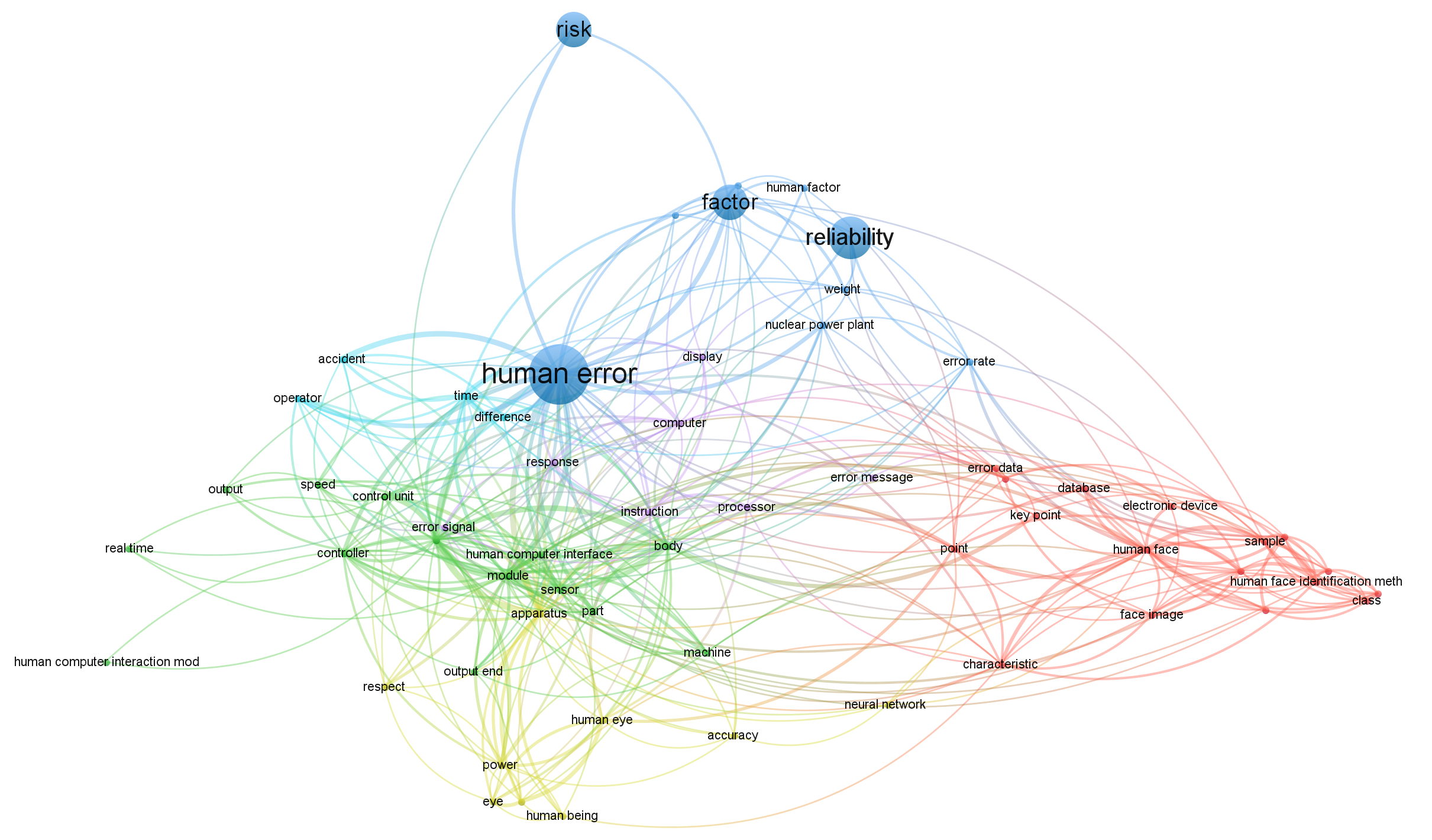}
\caption{Clustered Keyword Analysis of 'Human Error' Research Using VOSviewer}\label{pic_human_error}
\end{figure}

VOSviewer 1.6.17 is a software tool for constructing and visualizing bibliometric networks \cite{van2010software}. We collected 1,000 relevant indices on "human error" from the Web of Science, and the results from the software (Figure \ref{pic_human_error}) indicate a close association between human error and concepts such as risk and reliability. Additionally, there are applications where human error is managed as a control mechanism in real-time systems. The related research predominantly focused on human-computer interfaces, with some studies also examining individual characteristics, such as facial features and other human traits.

\subsection{Definition and Types of Human Error}
\label{Definition and Types of Human Error}

Human error refers to unintentional deviations from intended actions or decisions that can lead to adverse outcomes in safety-critical systems. These errors are rooted in cognitive processes such as attention, perception, memory, and decision-making, which directly impact human reliability. Reason \cite{reason2000human} distinguished between person-based causes—stemming from individual cognitive or emotional states, and system-based causes, which involve latent conditions that predispose environments to failure or degrade system defenses. Unlike unpredictable active failures, latent conditions are identifiable and thus amenable to proactive risk mitigation. Recognizing that while human variability is inevitable, work environments can be systematically optimized, Reason’s system perspective addresses multiple levels, from individuals and teams to tasks and organizational structures. Within this framework, human-computer interaction (HCI) plays a critical role in minimizing error through design and usability improvements. The Swiss cheese model conceptualizes accident causation as the alignment of weaknesses across multiple defense layers—technical, procedural, and human—that, when penetrated, allow hazards to result in failure. Further, Reason \cite{reason1990human} classified errors based on cognitive levels—skill-based, rule-based, and knowledge-based—and categorized them into mistakes, lapses, and slips, with the former two arising during planning and the latter during execution. This classification has been widely adopted in subsequent research on human reliability and error prevention \cite{lin2010optimizing, hobbs2002skills}.

\begin{figure}[h]
\centering
\includegraphics[width=0.9\textwidth]{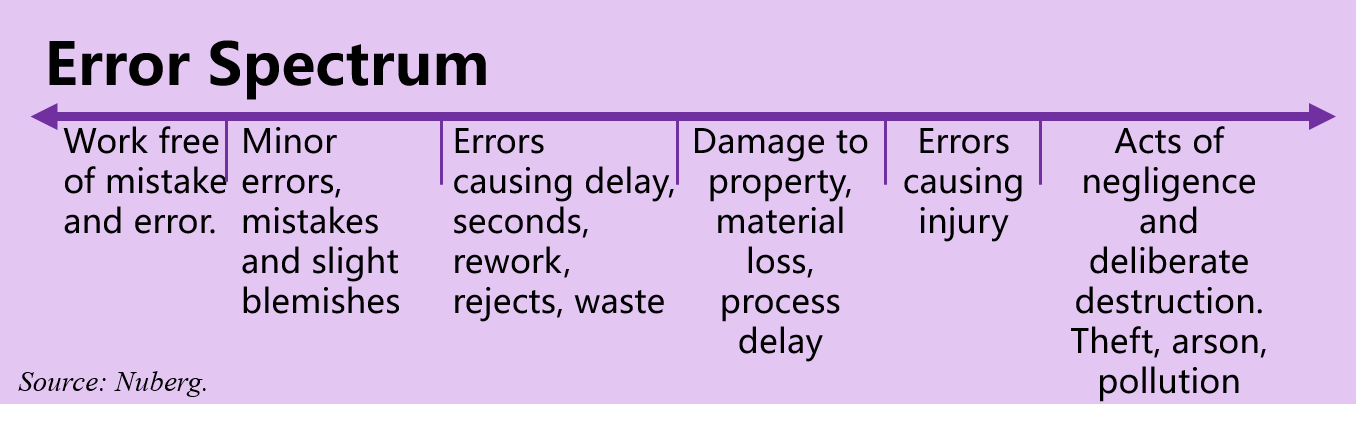}
\caption{Spectrum of Error Outcomes in Human Reliability Analysis}\label{spectrum}
\end{figure}

Since the early conceptualizations of human error, scholars have proposed various definitions and classifications. Petersen \cite{petersen2003human} broadly defined human error as any significant deviation from an established, required, or expected standard of human performance. Swain \cite{swain1963method} distinguished between two primary types: errors of commission (EOC), involving incorrect or unnecessary actions, and errors of omission (EOO), referring to failures to perform required actions. Kletz \cite{kletz2018engineer} further categorized errors into four types: slips or aberrations (e.g., operating pressurized equipment), errors preventable through improved training or instructions, errors stemming from physical limitations, and those resulting from excessive cognitive demands. Additionally, Nuberg \cite{petersen2003human} introduced a visual model illustrating the spectrum of error outcomes, offering insights into the underlying mechanisms and severity of human error (Figure \ref{spectrum}).

In modern engineering, human errors in HRA are typically classified into Types A, B, and C, with the latter further subdivided \cite{sharit2012human}, reflecting cognitive science insights that attribute errors to overload, intentional decisions, or workplace-induced traps \cite{petersen2003human}. Beyond general taxonomies, field-specific frameworks offer tailored perspectives—for example, the aerospace industry categorizes errors across cognitive, ergonomic, aeromedical, psychosocial, and organizational dimensions \cite{wiegmann2001human}, while safety management literature distinguishes mechanistic, individual, interactionist, and systems perspectives \cite{read2021state}. Team-related errors have gained prominence in high-risk domains, where factors like coordination and familiarity significantly impact performance \cite{sasou1999team, sieweke2015impact}. Programs such as TEAMSTEPPS \cite{alonso2006reducing} promote team safety culture by encouraging open communication and proactive error management \cite{helmreich2017culture, pronovost2003evaluation}. Organizational culture—defined by shared values and behaviors \cite{ouchi1985organizational, health1993organizing}—plays a critical role in shaping error responses. Although no universal definition of human error exists, the focus remains on practical strategies to manage and prevent errors effectively \cite{petersen2003human}.

\subsection{The Impact of Human Error on System Safety}
\label{The Impact of Human Error on System Safety}

Human error is a critical factor in compromising the safety of complex systems, especially in high-risk industries such as nuclear energy, aviation, and healthcare, where even minor mistakes can lead to severe consequences. Across safety-critical domains, it is widely acknowledged that human failure is responsible for the majority of accidents, ranging from 50\% to 90\%, depending on the industry (e.g., Baybutt 2002 \cite{baybutt2002layers}; Guo and Sun 2020 \cite{guo2020flight}; Shappell and Wiegmann 1996 \cite{wiegmann2001human}). For instance, the United States National Highway Traffic Safety Administration (NHTSA 2018 \cite{stewart2022overview}) attributes 94\% of road crashes to driver error, which has been interpreted as '94\% of serious accidents are caused by human error' (e.g., Rushe 2019 \cite{read2021state}).

Furthermore, numerous cases have demonstrated that human errors are a direct cause of accidents. In the nuclear energy sector, human error can lead to operational disruptions or, in the worst cases, nuclear accidents. For example, during the Three Mile Island accident on March 28, 1979 \cite{le2004human}, operator error, coupled with unclear control panel signals, resulted in a partial reactor meltdown. Similarly, in the Chernobyl disaster on April 26, 1986 \cite{stang1996chernobyl}, operators violated safety protocols and misjudged the reactor’s stability, leading to one of the most catastrophic nuclear accidents in history. In addition, many accidents in other fields have also been caused by human error. Notable examples include the challenger space shuttle disaster \cite{schwartz1987psychodynamics} on January 28, 1986, the JCO criticality accident in Tokaimura \cite{hayata2001cytogenetical}, Japan, on September 30, 1999, and the crash of a U.S. weather satellite\cite{fujita1977analysis} in November 1999.

These accidents have led to a widespread perception of a "human error problem," with solutions often focused on changing the people or their roles within the system \cite{woods2017behind}. For instance, some argue for reducing human involvement through increased automation, or for regulating human behavior with stricter monitoring, rules, or procedures. However, in practice, the issue is far more complex. Like the iceberg theory \cite{johnston1984hemingway}, the term "human error" is biased and tends to obscure the underlying causes of how systems truly function or fail. The real opportunity for constructive change and enhanced safety lies in examining the deeper systemic issues that are often masked by the human error label.

\subsection{Mitigation Strategies for Human Error}
\label{Mitigation Strategies for Human Error}

As discussed in section \ref{The Impact of Human Error on System Safety}, the issue of human error has garnered significant attention. Increasingly, scholars have been focusing on solutions within the system \cite{woods2017behind}. The error-recovery process may fall into any one of three stages: detection, indication, and correction \cite{love2004role}. Some scholars seek to replace human involvement with automation to avoid human errors, as discussed in detail in Section \ref{Artificial Intelligence (AI) in Human Error and Performance Analysis}. Other scholars have approached the issue from the perspective of human factors at the individual level, proposing wider-ranging management approaches. For example, Frank L. Greitzer et al. \cite{greitzer2014unintentional} provided strategic recommendations for mitigating unintentional insider threats. They focus on human factors and training, high-level organizational best practices, and automated defense mechanisms. Commonly used human factors prevention tools in industrial settings, such as self-checks, peer checks, and supervision, are also aimed at controlling individual errors. These tools are often derived from lessons learned through past incidents, as detailed in \cite{swain1983handbook}. Additionally, numerous mitigation strategies have been proposed to address the issue of unintentional information leakage by individuals \cite{wong2019human}. Wan Basri et al. \cite{ismail2018mitigation} also presented strategies for mitigating the risks associated with unintentional insider threats. They emphasize the importance of incorporating a risk-based approach, advocating for comprehensive risk assessments at every stage and reinforcing protective measures in areas identified as high-risk.

Furthermore, many scholars have focused on learning from accident cases to improve safety measures. For example, in response to the Fukushima accident, Don E. MacLeod et al. introduced a simplified human reliability analysis process for the deployment of emergency mitigation equipment (EME) \cite{macleod2014simplified}.  In addition, some scholars have emphasized the importance of improving human-machine interfaces \cite{lin2010optimizing, rasmussen1983skills, perez2011uncertainty, o2008human}, recognizing it as a critical area of research. For example, in 1997, Bob Fields et al. introduced THEA, a technique designed to help interactive system designers anticipate potential interaction failures or human errors once their designs become operational \cite{fields1997thea}. Broadly speaking, this falls under the domain of human-computer interaction (HCI) \cite{chignell2023evolution,fan2022human}, a highly popular research area. HCI exemplifies a systematic approach to controlling and reducing human errors by optimizing how humans interact with technology. Future optimization directions for human-machine interfaces can be guided by recent research \cite{amershi2019guidelines}.

However, these methods lack the ability to quantify the human error dynamically. As a result, risk-informed decision making was developed to dynamically mitigate the impact of human error. Further details on this topic can be found in Section \ref{Risk-Informed Decision Making and Human Reliability Assessment (HRA)}.

\section{Risk-Informed Decision Making and Human Reliability Assessment}
\label{Risk-Informed Decision Making and Human Reliability Assessment (HRA)}

Risk-informed decision making frameworks are essential for managing the complexity and uncertainty of safety-critical systems. By incorporating HRA into these frameworks, organizations can better understand the role of human error in overall system risk. This integration leads to more informed, effective, and resilient decision-making, ultimately enhancing both safety and operational performance in industries where human error can have severe consequences.

\subsection{Risk-Informed Decision Making Frameworks}
\label{Risk-Informed Decision Making Frameworks}

A key characteristic of risk is its inherent uncertainty. Rather than focusing solely on risk characterizations, the main emphasis is on the actions and decisions that need to be taken \cite{aven2019distinction}. Risk-informed decision-making (RIDM) is a strategic approach that integrates risk assessment into decision-making processes, especially in safety-critical industries. This approach is crucial because it enables organizations to balance safety with operational priorities such as cost, performance, and efficiency. By considering both the probability and consequences of potential failures, RIDM ensures that decisions are based on a comprehensive understanding of how different risks impact overall system safety.

We conducted a statistical analysis of documents retrieved from the Web of Science database, with results shown in Table \ref{tab1_risk}. Keywords such as "public goods game" appeared frequently, indicating that cooperation and resource allocation are core themes in risk-informed research. In contrast, terms like "centipede game" and "certainty effect" were less common. The prevalence of behavioral economics concepts suggests a strong reliance on this discipline to explain irrational decision-making under uncertainty and to support the development of behaviorally grounded models. The keyword distribution further implies that the field considers not only external risk factors but also internal cognitive biases and behavioral patterns as integral to decision-making. While classical theories such as the Allais and Ellsberg paradoxes are cited less often, their presence reflects ongoing efforts to validate and interpret irrational behaviors within uncertain contexts. Overall, risk-informed research demonstrates a multidisciplinary integration of behavioral economics, game theory, and psychology, with future efforts likely focusing on enhancing decision-making under uncertainty in cooperative and resource-constrained settings.

\begin{table}[h]
\caption{Frequency Analysis of Key Topics Related to 'Risk Informed' from Web of Science Literature}\label{tab1_risk}%
\begin{tabular}{p{7cm} p{5cm} }
\toprule
Key Word & Times  \\
\midrule
public goods game   & 835  \\
hyperbolic discounting   &525      \\
endowment effect    & 456    \\
demand characteristics  & 336     \\
decoy effect    & 324    \\
variety seeking   & 282     \\
reference dependence   & 277     \\
behavioral strategy & 233 \\
behavioral bias & 208 \\
behavioral game theory & 168 \\
Allais paradox & 168 \\
decision making under risk & 167 \\
Ellsberg paradox &139\\
\botrule
\end{tabular}
\end{table}

\begin{figure}[h]
\centering
\includegraphics[width=0.9\textwidth]{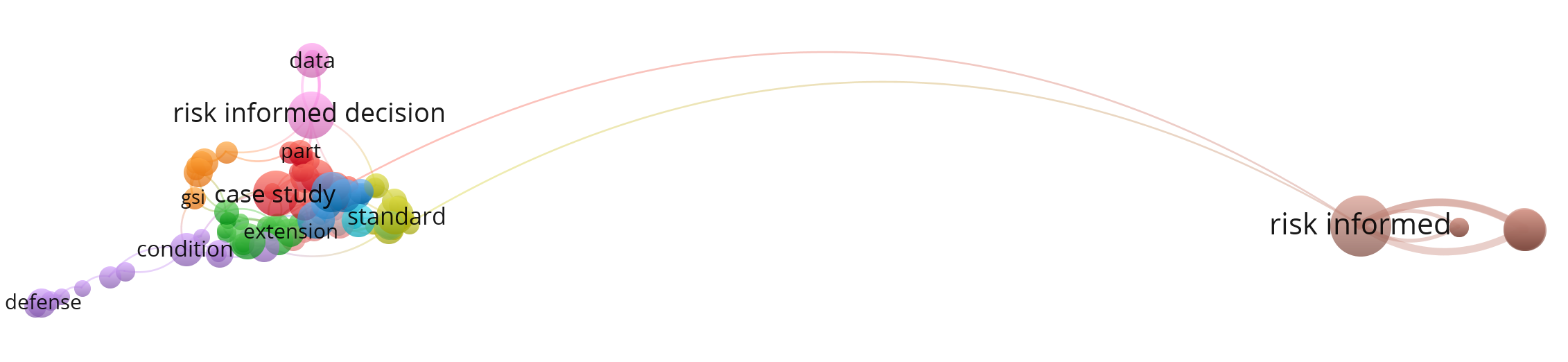}
\caption{Clustered Keyword Analysis of 'Risk Informed' Research Using VOSviewer}\label{pic_risk_informed}
\end{figure}

As shown in Figure \ref{pic_risk_informed}, research on the risk-informed approach remains relatively limited, as it is still a relatively new concept. Typically, risk-informed methods are integrated with decision-making processes and often rely on case studies to demonstrate the entire procedure. Additionally, this approach is closely related to safety regulation, loss control, and risk analysis. Current research in this area predominantly focuses on applications within the nuclear reactor sector.

Risk analysis involves identifying and managing situations that may cause harm to individuals, communities, or ecosystems, often through the interplay of human activities and natural processes \cite{amendola2002recent}. While decision-making incorporates algorithmic analysis, it also relies on inherently human elements such as intuition, emotion, and value judgments \cite{gondocs2024ai}. Traditional emergency decisions often lack quantitative rigor, prompting a shift toward a more structured and informed approach, as advocated by the US National Research Council \cite{stern1996understanding, xiao2024emergency}. This risk-informed paradigm has been widely applied across high-risk domains, including aerospace \cite{stamatelatos2006proposed}, nuclear power \cite{reinert2006including}, and industrial operations. It has also been extended to everyday contexts, such as school selection, where both academic and environmental risk factors are evaluated to support informed decision-making \cite{nakum2022developing} providing a structured approach for contemporary syllabus-based school selection.

\subsection{Human Reliability Assessment}
\label{Human Reliability Assessment}

Human Reliability Analysis (HRA) is essential for quantifying human error probabilities (HEPs) and assessing their impact on system reliability in complex, safety-critical environments. HRA methodologies have evolved from first-generation statistical approaches to second-generation methods emphasizing cognitive processes, and more recently to third-generation and dynamic models that incorporate real-time behavioral and cognitive changes. A statistical analysis of Web of Science documents (Table \ref{tab1_HRA}) reveals strong research emphasis on human-automation interaction, particularly in the automotive sector, where collaboration in manufacturing and system design is critical. The frequent occurrence of “systems engineering” underscores the integration of human reliability with system-level optimization across domains such as production, transportation, and aviation. Additionally, the association of “human error” with “root cause” highlights a focus on error diagnosis and mitigation to enhance system robustness. Although terms like “aircraft maintenance” and “safety engineering” appear less frequently, they indicate sector-specific concern for human reliability in safety-critical operations. Overall, HRA research trends reflect a growing emphasis on systems integration, automation, and AI to improve human reliability and minimize error-induced disruptions.

\begin{table}[h]
\caption{Frequency Analysis of Key Topics Related to 'Human Reliability' from Web of Science Literature}\label{tab1_HRA}%
\begin{tabular}{p{7cm} p{5cm} }
\toprule
Key Word & Times  \\
\midrule
automotive industry   & 26,126   \\
systems engineering   &10,219      \\
performance improvement    & 4,328    \\
FMEA   & 3,897     \\
human error    & 2,852    \\
business process automation   & 1,890     \\
design tool   & 1,513     \\
human reliability & 1,351 \\
root cause & 977 \\
safety engineering & 943 \\
aircraft maintenance & 906 \\
downtime & 881 \\
system safety & 582\\
\botrule
\end{tabular}
\end{table}

\begin{figure}[h]
\centering
\includegraphics[width=0.9\textwidth]{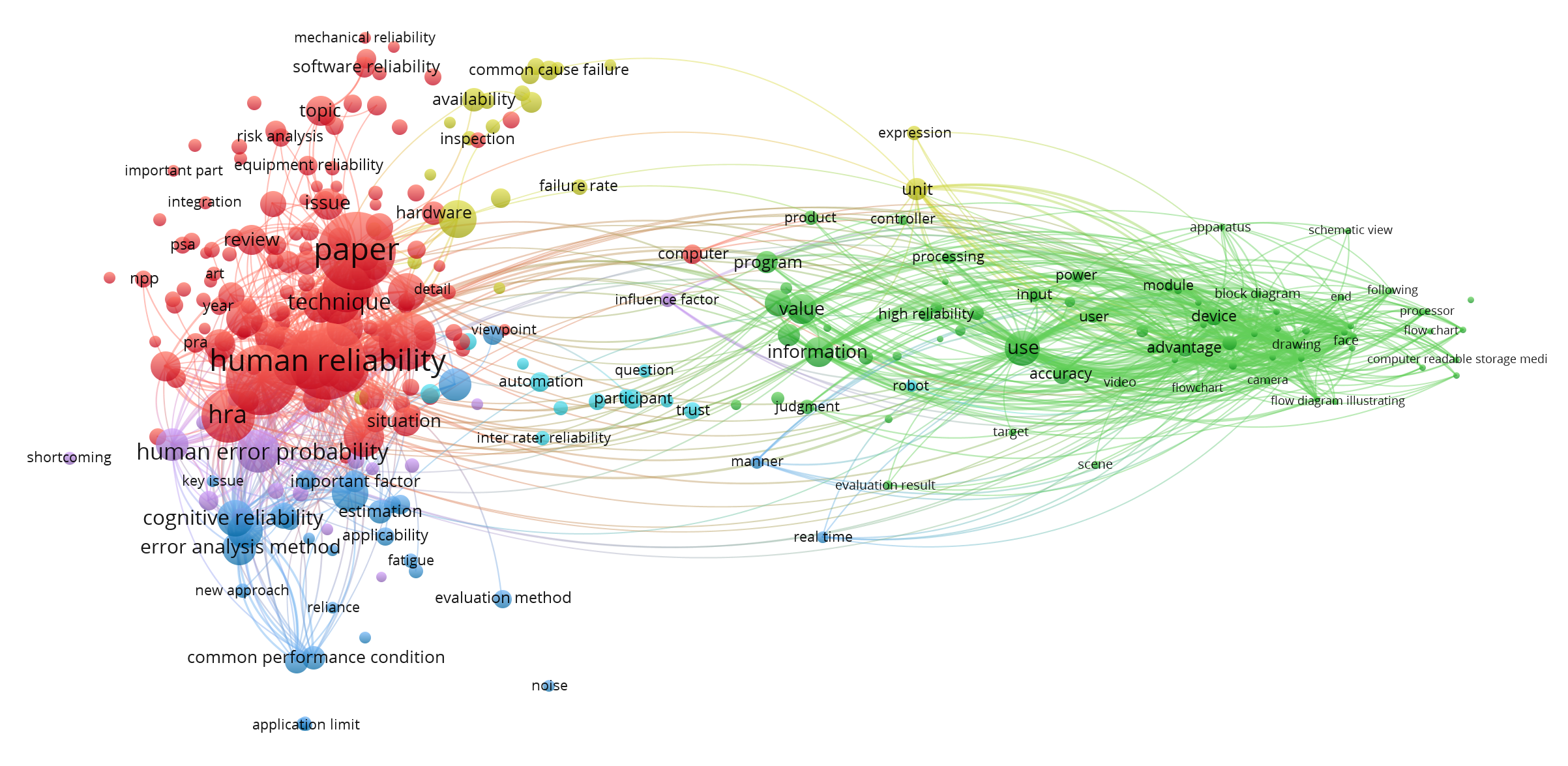}
\caption{Clustered Keyword Analysis of 'Human Reliability' Research Using VOSviewer}\label{human reliability}
\end{figure}

Using VOSviewer 1.6.17, we conducted a statistical analysis of documents from the Web of Science database. As shown in Figure \ref{human reliability}, research linking human reliability with cognitive reliability has gained prominence. The integration of human and system reliability, highlighted in green, is primarily associated with back-end applications. These studies emphasize detection and accuracy, with frequent implementations in nuclear power, human resources, and healthcare. Application-specific context is also carefully considered to ensure methodological relevance and suitability.

The first generation of Human Reliability Analysis (HRA) methods emerged between the 1960s and mid-1980s, laying the groundwork for quantifying human error in safety-critical systems. These methods—such as THERP, HCR, ASEP, SPAR-H, and EPRI HRA—relied heavily on expert judgment and statistical analysis grounded in observable operator behaviors, while largely neglecting underlying cognitive mechanisms. THERP \cite{swain1983handbook} provided a comprehensive framework through detailed task decomposition and fault tree modeling, offering point estimates for basic and conditional human error probabilities. However, its resource-intensive nature and limited cognitive insight posed challenges. EPRI HRA \cite{mccafferty1984annotated} offered a simpler, cost-effective alternative based on historical error data and performance shaping factors (PSFs), but similarly lacked depth in cognitive and team-level modeling. ASEP \cite{swain1983handbook} streamlined THERP for rapid assessments, though this came at the cost of reduced precision and static modeling. HCR \cite{swain1987accident} applied a Weibull distribution based on task completion times to model crew response failures, yet failed to incorporate dynamic or contextual factors. Despite their foundational contributions, first-generation HRA methods faced several limitations: inconsistent expert judgment, insufficient validation with simulator data, questionable psychological assumptions, and inadequate treatment of critical PSFs. Attempts to enhance these models have included integrating fuzzy logic \cite{castiglia2010risk, wang2018modified} and Bayesian theory \cite{bladh2008evaluation} to improve robustness and uncertainty quantification. Some studies, such as Castiglia et al. \cite{castiglia2015therp}, proposed hybrid approaches that combine methodologies like THERP and HEART to address these shortcomings. 

In addition to foundational first-generation HRA methods, several other approaches were developed in the 1980s to enhance task-level error analysis. SHARP \cite{suryoputro2017machinery} introduced a structured seven-step process that integrates empirical data to support quantitative assessment, but lacked cognitive and contextual modeling. SLIM-MAUD \cite{embrey1984slim} employed expert scoring and multi-attribute utility theory to calculate a Success Likelihood Index, offering flexible application across task types, yet was constrained by subjectivity and limited adaptability to dynamic contexts. HEART \cite{williams1988data} used predefined task classifications and error-producing conditions (EPCs) with adjustment factors to estimate error probabilities, capturing some real-world influences like fatigue and time pressure, though its reliance on user judgment and static modeling restricted broader applicability. SHARP1 \cite{hannaman1984systematic} extended SHARP through finer task decomposition and incorporation of error interdependencies via Error Tree Analysis (ETA), enhancing methodological precision but still lacking cognitive depth and dynamic responsiveness. Overall, while these methods advanced early HRA practice with improved quantification and PSF integration, they continued to face challenges in modeling cognition, psychological context, and operational variability.

The second generation of HRA methods emerged in the early 1990s. These methods focus on the dynamic cognitive processes involved in emergencies, such as detection, diagnosis, and decision-making, aiming to explore the mechanisms behind human errors. They combine cognitive reliability assessment with action execution reliability. Representative models include the CREAM, ATHEANA models. CREAM \cite{hollnagel1998cognitive} integrates task analysis with cognitive function failures and Common Performance Conditions (CPCs), supporting both qualitative and quantitative evaluations of human reliability. Its strength lies in modeling perception, judgment, and execution within operational contexts, though it is resource-intensive and requires specialized expertise. ATHEANA \cite{cooper1996technique} introduces Cognitive Error Mechanisms (CEMs) and error trees to trace scenario-based error progressions, making it well-suited for low-frequency, high-consequence events. However, its complexity and reliance on expert input limit routine applicability. SPAR-H \cite{gertman2005spar}, in contrast, offers a simplified and standardized approach using predefined task types, baseline human error probabilities, and eight performance shaping factors. While efficient for time-constrained assessments, SPAR-H lacks deep cognitive modeling and exhibits limited adaptability to complex, context-sensitive tasks. Collectively, these methods advance beyond first-generation approaches by embedding cognitive and contextual reasoning but vary significantly in analytical depth, interpretability, and practical feasibility.

Third-generation HRA emphasizes dynamic, context-sensitive modeling through methods such as IDHEAS-G and Phoenix. IDHEAS-G incorporates 71 cognitive function modules and 20 PIF types, enabling flexible, task-specific modeling. Phoenix follows a qualitative four-step process to assess human failure events, later extended by PhoenixPRO, which integrates event sequence diagrams, fault trees, and Bayesian belief networks for application in the petroleum industry \cite{ekanem2016phoenix, ramos2020human}. Simulator-based tools have also advanced third-generation HRA. SHERPA and PROCOS analyze variations in human error across operational scenarios \cite{di2015simulator, leva2009quantitative}, while the IDAC model focuses on crew decision-making under dynamic conditions \cite{chang2007cognitive1, mosleh2004model}. In parallel, data-driven approaches like HERA \cite{hallbert2004using} leverage operational histories to inform HRA, using nine analytical indices including event diagnosis, system complexity, and dependency factors. Hybrid methods have also emerged, such as Maya et al.’s integration of hierarchical task analysis (HTA) with HEART to identify and mitigate errors in maritime emergency contexts \cite{de2022practical}. Collectively, these approaches reflect the third generation’s shift toward integrated, cognitively grounded, and simulation-supported human reliability assessment.

Recent advancements in Human Reliability Analysis (HRA) have focused on developing dynamic HRA (D-HRA) approaches that integrate classical methods with time-sensitive modeling techniques. For example, Liu et al. \cite{li2025improved} proposed a D-HRA framework for nuclear control rooms by combining SPAR-H with system dynamics to simulate the temporal evolution of performance shaping factors (PSFs). Similarly, tools like EMRALD \cite{park2025dynamic} have been employed to support dynamic risk-informed HRA. Other studies address operator behavior in stochastic and time-critical emergency scenarios or under extreme collaborative conditions, reflecting a broader shift toward incorporating system dynamics, agent-based modeling (ABM), cognitive modeling, and real-time monitoring into HRA. This evolution transforms HRA from static assessment into continuous, predictive analysis. While third-generation HRA is often characterized by dynamic features, it is more accurately described as dynamic risk-informed HRA, aligning with dual objectives: proactively identifying human error risks and retrospectively analyzing influencing factors. These aims are consistent with risk-informed system frameworks, as introduced by Siu and Collins \cite{siu2008research}, who advocated for integrating human error into risk-informed decision-making—a foundation for modern D-HRA.

\subsection{HRA in Risk-Informed Frameworks}

Human error is a significant contributor to system failures in safety-critical industries. To effectively control human error, some studies have integrated HRA techniques into risk-informed decision-making (RIDM) frameworks. For instance, Joe et al. combined the THERP introduced in Section \ref{Human Reliability Assessment} with human cognitive reliability. They integrated HRA into the RISMC framework to enable dynamic simulations of human reliability, which can be used in risk-informed safety margin characterization \cite{joe2015development}. Blackett proposed integrating HRA and HFE to enhance risk-informed plant design and improvement \cite{taylor2017integrating}. This integration seeks to improve plant safety by addressing human error through HRA and optimizing human performance via HFE-based design. By combining these two approaches, a more comprehensive, risk-based strategy for plant design is achieved, enhancing both safety and operational efficiency. Similarly, Deng et al. integrated HFE with PSA/HRA analysis to develop a comprehensive framework for risk-informed decision support \cite{shiguang2022research}.

Mechanistic approaches have also been explored, such as the performance evaluation of teamwork (PET) method proposed by Petkov \cite{petkov2017team}, which evaluates team performance using context-sensitive models of individual cognition and mutual communication. Built on probabilistic interpretations of mental processes and network reliability, PET integrates holographic-like behavior, objective–subjective perceptions, and a statistical representation of the human-organization-technology (HOT) system. It supports classical, Bayesian, and quantum probabilities, modeling macroscopic context through object, image, and situation components, offering a comprehensive foundation for further refinement.

While HRA techniques provide valuable insights into human reliability, they also face challenges in dealing with cognitive complexity, uncertainties, and data limitations. For HRA to be fully effective within risk-informed decision making frameworks, ongoing research, and improvements in data collection, cognitive modeling, and error prediction are essential. Notably, many experts have contributed efforts to address data limitations, with details available in \cite{bye2023future}.

\section{Cognitive Science and Human Performance Models}
\label{Cognitive Science and Human Performance Models}

Cognitive science provides essential insights into the mechanisms underlying human error and performance, especially in high-risk environments. By understanding the cognitive foundations of human behavior, we can develop more accurate models to predict and mitigate errors. This section explores the cognitive basis of human error, its application in performance models for high-risk systems, and the integration of cognitive science with HRA and AI to enhance decision-making and system reliability.

\subsection{Cognitive Foundations of Human Error}
\label{Cognitive Foundations of Human Error}

\begin{figure}[h]
\centering
\includegraphics[width=0.9\textwidth]{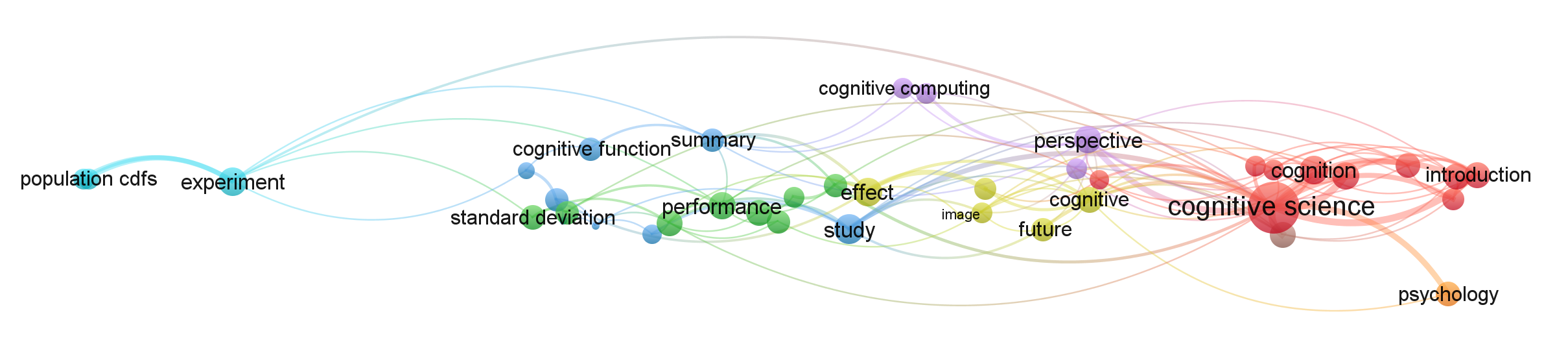}
\caption{Topic Map of Cognitive Science: A Statistical Analysis Based on Web of Science Publications}\label{pic_Cog}
\end{figure}

Cognitive science plays a crucial role in understanding the mechanisms behind human error by exploring how human cognitive processes—such as attention, memory, and decision-making—function and how they can lead to errors in complex systems \cite{alvarenga2019review}. As discussed in Section \ref{Definition and Types of Human Error}, many definitions and classifications of human error are based on principles from cognitive science. We conducted a statistical analysis using documents from the Web of Science database. As shown in Figure \ref{pic_Cog}, related research is often closely integrated with psychology and frequently involves experimental studies. Additionally, there are intersections with neuroscience, as well as considerations of religious factors and the role of cognition within group dynamics.

\subsection{Human Performance Models in High-Risk Systems}
\label{Human Performance Models in High-Risk Systems}

Human performance models (HPMs) aim to understand, simulate, and predict behavior in complex, high-stakes environments by integrating theories from cognitive science, psychology, and behavioral analysis. A statistical analysis of Web of Science documents (Table~\ref{tab1_HPM}) reveals that HPM research heavily utilizes simulated environments, particularly driving simulators, to study cognitive and behavioral responses under controlled conditions. Increasingly, cognitive science and AI are converging in this field, as indicated by keywords like 'cognitive architecture' and 'embodied agents,' reflecting a shift toward modeling internal cognitive mechanisms. Application-driven research dominates, with frequent references to domains such as aviation and task optimization. Emerging interest in 'human systems integration' and 'ecological interface design' signals a trend toward aligning performance modeling with system design. Overall, HPM research is evolving through interdisciplinary collaboration, expanding its scope to support performance optimization and safety in complex systems.

\begin{table}[h]
\caption{Frequency Analysis of Key Topics Related to 'Human Performance' from Web of Science Literature}\label{tab1_HPM}%
\begin{tabular}{p{7cm} p{5cm} }
\toprule
Key Word & Times  \\
\midrule
driving simulator   & 5,498  \\
test design   &692       \\
virtual humans   & 668    \\
embodied agent  &642    \\
air traffic controller   & 564     \\
cognitive architectures   & 553    \\
inspection time   & 551     \\
cursor & 476 \\
new learning & 399 \\
cognitive ergonomics & 364 \\
cognitive engineering & 364 \\
tasking & 234 \\
ecological interface design & 199 \\
augmented cognition & 181 \\
automatic and controlled processes & 171 \\
work domain analysis & 132 \\
human performance modeling & 123 \\
human systems integration & 85 \\
\botrule
\end{tabular}
\end{table}

\begin{figure}[h]
\centering
\includegraphics[width=0.9\textwidth]{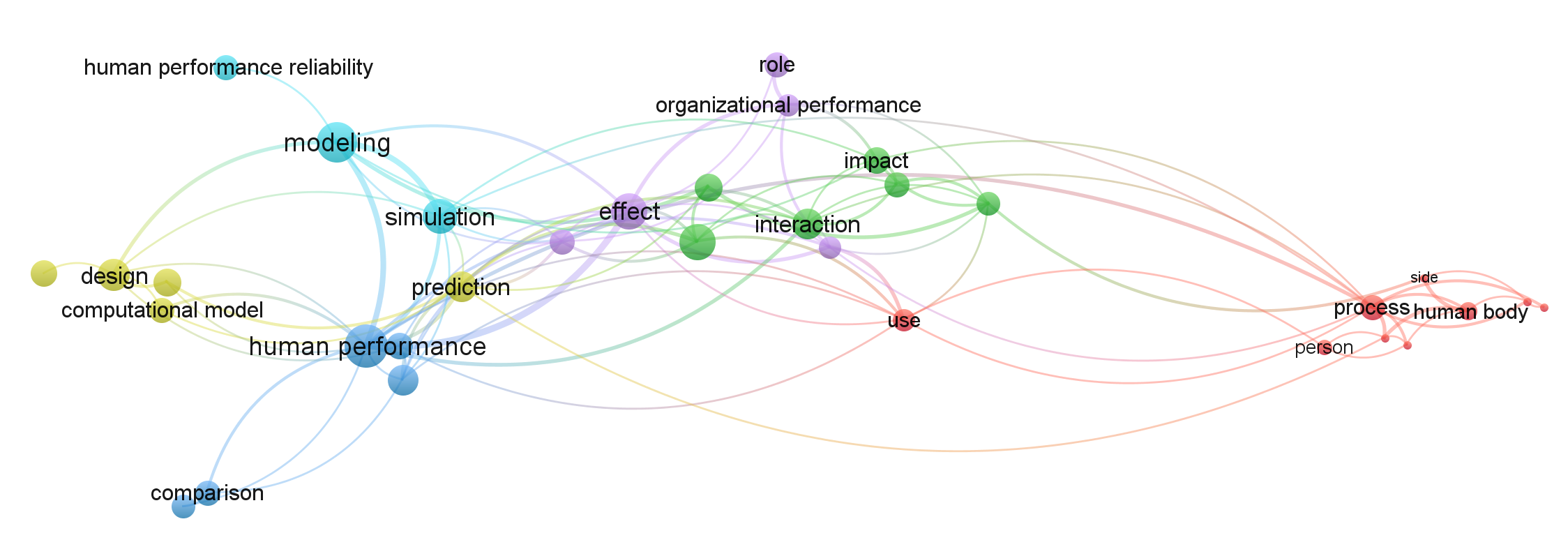}
\caption{Clustered Keyword Analysis of 'Human Performance Model' Research Using VOSviewer}\label{pic_HPM}
\end{figure}

As shown in Figure \ref{pic_HPM}, research on HPMs primarily focuses on simulations and the development of human models. Additionally, it often considers the impact of HPM on interactions within organizational groups. Case studies are commonly used to demonstrate the application of these models, and some research also integrates aspects of human physiology. This field is particularly concentrated in the area of human resources, where understanding and optimizing human performance is critical.

Li et al. \cite{li2020human} summarized that the research field of HPM encompasses four aspects. The first is physical models implemented in software, such as Jack \cite{blanchonette2010jack}, Safework \cite{fritzsche2010ergonomics}, and HUMOSIM \cite{reed2006humosim}, which include components like anthropometry, biomechanics, kinematics, physiology modeling, and simulation. The second aspect involves models based on control theory \cite{jagacinski2018control}, representing a powerful and dynamic part of HPM. The third includes models grounded in cognitive architectures, such as ATC-R \cite{anderson2004integrated}, where memory, attention, perception, and action are integrated into software. The final aspect is part-task models designed to predict excessive physical stress, with the potential to be integrated with ergonomic models. Due to space constraints, this paper will focus solely on the models grounded in cognitive architectures, while other aspects can be referred to in ref \cite{li2020human}.

With advances in cognitive science, HPMs increasingly incorporate elements such as perception, memory, learning, problem-solving, and decision-making, providing a more comprehensive simulation of human performance. Influenced by Colonel John Boyd’s OODA model—perception, cognition, decision, and action—HPMs now often align with neuroscience, mapping cognitive functions to brain regions and validating findings through EEG studies \cite{boyd2018discourse, anderson2008central, cao2013queueing}. These models deepen understanding of cognition and inform practical applications in interface design and human-machine function allocation \cite{yuan2021human}.

In human-computer interaction, researchers have developed models to predict human performance during task execution by drawing analogies between humans and machines—both process inputs and outputs and are influenced by environmental factors \cite{deneulin2009introduction}. Human performance modeling (HPM) employs mathematical or computational methods to simulate task-related information processing \cite{pew2008more}. Foundational models such as Fitts’ Law \cite{fitts1954information} and Hick’s Law \cite{hick1952rate}, rooted in information theory, address specific performance aspects but fall short of simulating holistic behavior. More comprehensive cognitive models—including SOAR, EPIC, ACT-R, and QN-MHP—offer diverse frameworks for modeling cognition and performance.SOAR \cite{laird1987soar} employs chunking and reinforcement learning to model procedural knowledge and learning curves but struggles with multitasking and realistic stimulus processing \cite{cooper1995soar}. EPIC \cite{kieras1996epic} features modular perception, motor, and memory components operating in parallel with stochastic timing, yet its limited capacity to simulate human learning and memory has restricted its development \cite{kieras1997overview}. ACT-R \cite{anderson2004integrated} integrates symbolic and perceptual-motor modules within a unified framework and has seen widespread application across domains such as education, driving, aerospace, and nuclear power \cite{ritter2019act}, though it is constrained by its sequential production system, limiting multitasking simulation. QN-MHP \cite{liu2006queueing}, grounded in queuing theory, excels at modeling multitask human performance with strong scalability and quantitative output \cite{zhang2022cognitive}, but its cognitive abstraction and context insensitivity limit its explanatory depth. Collectively, these models contribute valuable insights into human performance, yet trade-offs remain between fidelity, scalability, and cognitive realism.

\subsection{Integrating Cognitive Science with HRA and AI}
\label{Integrating Cognitive Science with HRA and AI}

Cognitive science offers a robust framework for understanding human error by examining attention, memory, and decision-making. Studies have applied cognitive theories to refine error classification and improve human performance analysis. Wahr et al. reclassified human errors into cognitive-, system-, and communication-based categories, building on Reason’s taxonomy \cite{reason1990human, wahr2024human}. The U.S. NRC synthesized insights from cognitive and behavioral sciences into a framework of five macrocognitive functions: detection, sensemaking, decision-making, action, and teamwork \cite{whaley2016cognitive}. This foundation is further operationalized in IDHEAS-G and IDHEAS-ECA through structured cognitive components for systematic error analysis.

Integrating psychological methods into HRA to analyze the mechanisms of human error is a key trend for future development. For example, Akyuz et al. \cite{akyuz2014utilisation} utilized the human factors analysis and classification system (HFACS) combined with the cognitive map (CM) technique to model human error in the context of marine accident analysis and prevention. Plant et al. \cite{plant2013explanatory} discussed the future of schema theory, highlighting its potential role as a unifying framework in ergonomics and its contribution to our understanding of distributed cognition. The advantage of second-generation HRA methods over the first lies in their consideration of human factors. An example is the ATHEANA introduced in Section \ref{Human Reliability Assessment}, it is a multidisciplinary HRA framework that introduces a general, albeit preliminary, process for defining human failure events and estimating their probabilities through iterative search schemes. It also features a knowledge base that links PSFs to unsafe actions \cite{cooper1996technique}.

Performance influencing factors (PIFs), often grounded in cognitive science, are central to human reliability qualification. For example, a workshop identified key PIFs for three cognitive failure modes: forgetting to transmit, transmission error, and reception error \cite{isaac2002human}. In practice, human performance is frequently used as a proxy for reliability, yet mechanistic studies linking performance metrics to human reliability remain limited. This gap underscores the need for research that systematically maps performance to reliability, especially in complex operational contexts.

Digital twin (DT) technology, which enables accurate digital representations of physical entities, has gained significant attention across academia and industry. While widely applied in domains like smart manufacturing and transportation, traditional DTs typically model non-living systems. To extend this concept to human-centric applications, the human digital twin (HDT) has emerged, offering dynamic in silico representations of physiological, psychological, and behavioral states. HDTs hold great promise for personalized healthcare, supporting remote monitoring, diagnosis, treatment, and rehabilitation. Despite their potential, HDTs face substantial research challenges and have become a growing focus of study \cite{chen2023networking}.

\section{AI in Human Error and Performance Analysis}
\label{Artificial Intelligence (AI) in Human Error and Performance Analysis}
Integrating AI into human error and performance analysis enhances the ability to predict, detect, and assess reliability. This approach is expected to become a significant trend in the future. AI complements traditional methods by identifying error patterns and improving decision-making. This section discusses AI’s role in predicting human error, enhancing HRA, and modeling human performance.

\subsection{Introduction to AI}

The term Artificial Intelligence was first introduced by John McCarthy in the mid-1950s \cite{mccarthy1978history}, reflecting early efforts to replicate human cognitive functions such as pattern recognition, reasoning, and learning. Today, AI plays a pivotal role in real-time monitoring, prediction, and analysis of human behavior, enabling more accurate error detection and prevention. A statistical analysis of Web of Science documents (Table \ref{tab1_ai}) indicates that while terms like 'deployment,' 'weighting,' and 'multidimensional analysis' are less frequent, the dominant focus remains on data acquisition, processing, and quality control. Mobile platforms—particularly smartphones—have become central to AI deployment in areas such as health monitoring and augmented reality. The frequent appearance of terms like 'open source' and 'benchmarking' further highlights the importance of collaborative tools and standardized practices. Collectively, these trends signal AI’s evolution from theoretical development to practical, data-driven applications supported by mobile technologies and open ecosystems.

\begin{table}[h]
\caption{Frequency Analysis of Key Topics Related to 'Artificial Intelligence' from Web of Science Literature}\label{tab1_ai}%
\begin{tabular}{p{7cm} p{5cm} }
\toprule
Key Word & Times  \\
\midrule
artificial intelligence data analysis   &  105,874  \\
smartphone   &42,002       \\
mobile phone    & 32,777    \\
mobile device  & 32,695    \\
data collection    & 20,762     \\
benchmarking   & 18,121    \\
open source   & 15,292     \\
data quality & 12,227 \\
mobile app development & 11,647 \\
\botrule
\end{tabular}
\end{table}

\begin{figure}[h]
\centering
\includegraphics[width=0.9\textwidth]{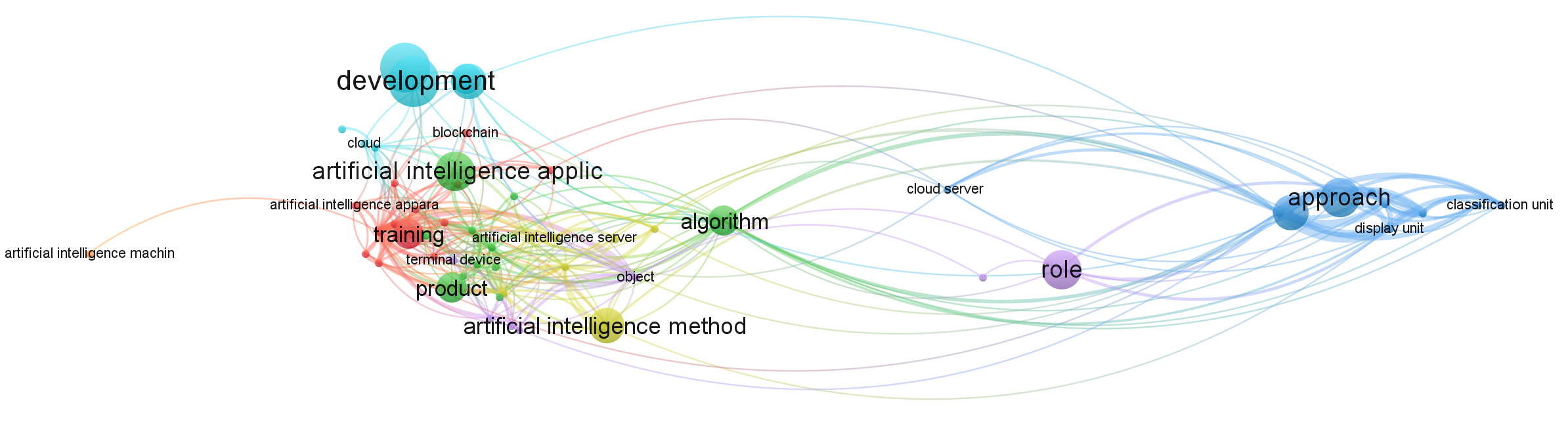}
\caption{Clustered Keyword Analysis of 'Artificial Intelligence' Research Using VOSviewer}\label{pic_AI}
\end{figure}

As illustrated in Figure \ref{pic_AI}, current AI research primarily centers on training, algorithms, data, and application areas such as classification and prediction. Historically, AI has evolved through four major paradigms: symbolic, Bayesian, connectionist, and evolutionary. The symbolic school conceptualizes cognition as symbolic manipulation, foundational to traditional methods like HRA and fault tree analysis \cite{udrescu2020ai, ruijters2015fault}. The Bayesian approach models uncertainty using conditional probabilities and Bayes’ theorem, enabling dynamic inference as new evidence emerges \cite{van2021bayesian}. The connectionist school draws inspiration from the human nervous system, modeling intelligence through interconnected artificial neurons. This paradigm underlies modern neural network architectures, including MLPs, CNNs, RNNs, and Transformers \cite{bishop1994neural, kruse2022multi, li2021survey, medsker2001recurrent, vaswani2017attention}. Each school offers a distinct computational framework for learning and reasoning, contributing to the diverse methodological landscape of contemporary AI.

The evolutionary school views intelligence as a product of biological evolution, simulated through algorithms incorporating genetic variation, crossover, and selection. These evolutionary algorithms serve as model-agnostic optimization tools, applicable to symbolic systems, probabilistic models, and neural networks, particularly in complex or ill-structured environments \cite{kaelbling1996reinforcement}. With the rise of large language models (LLMs), AI is increasingly categorized into general knowledge-based and expert knowledge-based systems. While domain-specific models address specialized tasks \cite{thirunavukarasu2023large, kaddour2023challenges}, general models aim to support broader reasoning and computation capabilities \cite{chang2024survey}. In human error analysis, virtual agents trained with both expert and general knowledge offer a hybrid approach to error simulation and mitigation. Technological advances have further driven convergence among cognitive paradigms—symbolic tools like knowledge graphs are now integrated with neural and Bayesian models, and rule-extraction techniques enhance neural interpretability, reflecting symbolic principles. This methodological fusion supports more robust and explainable AI, with Bayesian methods remaining particularly influential in reliability engineering \cite{weber2012overview}.

\subsection{AI for Human Error Prediction and Detection}

Advances in AI have made it increasingly applicable to supplement expert’s decision-making in the form of a decision support system on various tasks \cite{lee2021human}. In the realm of prediction, research primarily focuses not on predicting human errors directly but on providing decision-support systems aimed at reducing such errors. For example, Göndöcs et al. \cite{gondocs2024ai} explored the application of decision support in the medical field, discussing how AI-driven systems can assist in minimizing human error during critical decision-making processes.  Some research has increasingly focused on fault diagnosis to avoid human errors in critical tasks. For instance, Karissa C. Hammer et al. \cite{hammer2022using} used AI to improve the accuracy of embryo identification, achieving a 100\% success rate in their case studies. In the nuclear industry, Gursel et al. \cite{gursel2023using} developed an unsupervised anomaly detection technique based on generative adversarial networks (GANs) to detect anomalies in manually collected surveillance data from nuclear power plants. Similarly, Morais et al. \cite{morais2022identification} proposed an automated approach for classifying accident reports, reducing processing time, improving efficiency, and alleviating human workload. Furthermore, Parhizkar et al. \cite{parhizkar2021supervised} introduced a method to predict the most probable scenarios with acceptable accuracy in a very short time, demonstrating the potential of AI in optimizing decision-making processes. 
Emilie D et al. proposed a partially observable markov decision process (POMDP) model, leveraging sensor data for action planning and human error detection during activities of daily living (ADLs) \cite{jean2015pomdp}. Similarly, Lee et al. found that after tuning an AI assistant system based on therapist feedback, its performance improved significantly, with average F1-scores increasing from 0.8377 to 0.9116 (p < 0.01) \cite{lee2021human}. The increasing intelligence of machines leads to a shift from HCI to human-machine cooperation (HMC)\cite{hoc2001towards, van2005integrating, zhang2021ideal}.

However, their work primarily relies on traditional AI methods. In contrast, AGI (Artificial General Intelligence) systems are expected to be more broadly focused, equaling or surpassing human intelligence across a wide range of cognitive abilities \cite{everitt2018agi, gurkaynak2016stifling}. AGIs will be capable of autonomous planning, reasoning, decision-making, and problem-solving, even for tasks they were not originally designed to handle \cite{kaplan2019siri}. Paul M. Salmon et al. identified fifteen categories of human factors and ergonomics (HFE) methods and explored their potential role in AGI system design. These HFE methods include task analysis, human error identification, situation awareness analysis, and mental workload assessment, among others \cite{salmon2021putting}. The integration of such methods not only facilitates the implementation of AGI but also helps to reduce human error.

\subsection{AI-Enhanced Human Reliability Assessment}

AI offers significant enhancements to traditional HRA by enabling automated data processing \cite{schelter2018automating}, improving error detection \cite{gehring1993neural}, and facilitating real-time performance monitoring \cite{robertson2014review}. Unlike static models, AI allows for dynamic updating of human error probabilities in response to changing cognitive loads, task complexity, or environmental stressors. Many recent approaches integrate Bayesian networks to support probabilistic reasoning. For instance, Zhang et al. \cite{zhang2020probabilistic} combined THERP with Bayesian modeling and sensitivity analysis to inform operator training, while Cai et al. \cite{cai2013dynamic} applied dynamic Bayesian networks for offshore blowout risk assessment. In marine contexts, Yang et al. introduced both a modified CREAM-Bayesian framework \cite{yang2013modified} and a fuzzy rule-based Bayesian model for FMEA \cite{yang2008fuzzy}. Li et al. \cite{li2022reliability} extended this to multi-state system reliability under common cause failures using fuzzy probabilities, and Zarei et al. \cite{zarei2019hybrid} proposed a hybrid model integrating HFACS, intuitionistic fuzzy theory, and Bayesian networks. While deep learning has been applied to enhance equipment reliability \cite{chen2020time, xiao2023enhancing, qi2024multimodal}, its application to improving human reliability remains limited and underexplored.

On the other hand, with the emergence of large language models (LLMs), numerous studies have explored their potential to simulate human behavior. For instance, Xie et al. \cite{xie2024can} developed interactive agents and discovered that LLM agents, particularly GPT-4, exhibit high behavioral alignment with humans in trust-related behaviors, suggesting the feasibility of simulating human trust dynamics with LLM agents. Li et al. \cite{li2024econagent} introduced EconAgent, a novel intelligent agent based on LLMs, which incorporates perception, memory, and action modules to simulate individual behaviors. Extensive experiments across various macroeconomic scenarios demonstrated that EconAgent performs effectively in modeling macroeconomic activities. Additionally, Zhang et al. \cite{zhang2023exploring} investigated cooperation mechanisms among LLMs in a multi-agent social framework. By constructing a multi-agent simulation, the study examined how LLM agents with different traits and cognitive styles cooperate. The results showed that certain cooperative strategies were more effective than previous leading methods, and LLM agents exhibited human-like social behaviors. Furthermore, the study integrated insights from social psychology to better understand these agents' social interactions. Overall, LLMs have shown great promise in simulating scenarios that require interaction and collaboration \cite{bansal2021does}, although the integration of LLMs in such contexts is still relatively rare.

It is important to note that AI is not without its challenges. Many scholars have expressed concerns about AI, such as Mark Ryan, who opposes anthropomorphizing AI, as it can lead to issues of accountability \cite{ryan2020ai}. Additionally, Choudhury et al. \cite{choudhury2022impact} found that factors like performance expectancy, perceived risk, and trustworthiness influence practitioners' perceptions of AI's impact on decision-making.

\subsection{AI and Human Performance Models}

The integration of AI with human performance models (HPMs) facilitates the development of real-time decision support systems that are dynamic and personalized, adapting to an operator’s cognitive and behavioral state. Currently, the integration of AI with HPMs is expanding across various domains, including human-AI collaboration \cite{wang2020human}, human-machine interfaces \cite{heng2022flexible}, driver assistance systems \cite{bengler2014three}, virtual human technology \cite{chen2023diffusiontalker}, and cognitive workload analysis. However, most of the existing work focuses on detection, diagnosis, and optimization, with a lack of mechanistic understanding. For example, in human performance monitoring, physiological signals such as electroencephalogram (EEG) are commonly used \cite{steinhauser2010decision}, and virtual reality (VR) is often employed to simulate real-world scenarios \cite{morbi2012reducing}. However, the analysis often remains superficial, without delving into the underlying mechanisms. 
In addition to these physiological signal-based approaches, some models operate on a more macro level.  For example, ALBAYRAK and ERENSAL utilized the analytic hierarchy process (AHP) to enhance human performance by addressing multiple criteria decision-making problems \cite{albayrak2004using}. Their model takes a broader perspective, considering factors such as the physical workplace, organization of work, human performance capabilities, attitudes, and managerial factors, providing a comprehensive framework for understanding human performance from a higher-level viewpoint.

\section{Current Challenges and Future Research Directions}
\label{Current Challenges and Future Research Directions}

Despite significant advancements in HRA, AI, and cognitive science, there remain several challenges in integrating these technologies to fully address human error in high-risk environments. The specific challenges and future directions are summarized as follows:

\begin{itemize}
    \item \textbf{Limited availability of human reliability data.} A major challenge is the lack of high-quality, real-time data on human performance, which constrains the training and validation of AI models. Addressing this requires not only algorithm development but also standardized data collection and processing frameworks. Enhanced collaboration among academia, industry, and regulators is essential to establish data-sharing protocols and privacy standards. Bridging theory and practice through grounded theory and operator interviews can further enrich datasets with real-world insights \cite{goldfarb2021prediction, oktay2012grounded}.

    \item \textbf{Insufficient application of AI in current practices.} Despite AI's growing presence, trust in its ability to model human behavior remains low \cite{lenskjold2023should}. Critics argue that algorithms struggle to capture human complexity \cite{barassi2021human}. While AI is seen as the future, building trust requires technical advances, transparency, and ethical safeguards. Currently, AI is mainly applied to equipment reliability, with limited use in modeling or assessing human performance. Future research should focus on developing AI systems capable of accurately simulating human behavior and interactions.

    \item \textbf{Reliance on subjective knowledge in human modeling, Simulation, and Reliability Assessment.} Current approaches to human modeling and HRA often rely on subjective judgment. Advancing the field requires integrated frameworks that draw on cognitive science, engineering, AI, and human factors \cite{asan2021research, chignell2023evolution}. Interdisciplinary collaboration can foster more objective, data-driven models, enhancing safety and reliability in human-critical systems.

    \item \textbf{Lack of resilient mechanisms for quantitative analysis of human error.} Traditional HRA emphasizes error prevention but often overlooks human resilience—the ability to adapt under stress and maintain performance \cite{besnard2003cognitive}. A resilience-oriented approach would consider factors like fatigue, personal stressors, and unexpected events. Integrating resilience into HRA, especially through AI, can enhance both error prediction and adaptive capacity, leading to more robust and supportive systems.
    \item \textbf{Lack of rigorous quantitative approaches from cognitive models to human reliability models.} A gap remains in mechanistic research connecting human performance metrics to reliability outcomes. Developing frameworks that quantitatively link factors like decision-making and adaptability to system reliability is essential for more accurate and comprehensive HRA.
\end{itemize}

\section{Conclusion}
\label{Conclusion}

Human error continues to play a pivotal role in the safety and reliability of high-risk industries, where even minor mistakes can lead to severe consequences. This review has demonstrated the importance of incorporating human error analysis within risk-informed frameworks by integrating AI, HRA, cognitive models, and HPMs. It begins by presenting an in-depth exploration of human error in high-risk environments, examining its definition, types, and the significant impact it has on system safety. It also discusses various mitigation strategies to address these errors, aiming to enhance overall safety and performance. Secondly, it transitions to risk-informed decision-making frameworks, focusing on HRA and how these assessments play a critical role in managing risks within complex systems. The integration of HRA with risk-informed approaches is highlighted, providing a clearer understanding of its application in decision-making processes. Then, further analysis is dedicated to cognitive science and HPM, where the cognitive foundations of human error are explored. This includes an examination of how these models function within high-risk systems, and how cognitive science can be integrated with HRA and AI to enhance understanding and predictability. The role of AI in human error prediction, detection, and AI-enhanced HRA is also discussed.

Despite significant advances in the integration of HRA, AI, and HPMs, there remain numerous challenges. These include issues such as data scarcity, insufficient application of AI algorithms, and an over-reliance on subjective expert knowledge. To address these challenges, it is essential to bridge the gap between academic research and real-world applications. One promising approach is to utilize grounded theory and conduct interviews with operators to collect more comprehensive and practical data. Additionally, we propose that human factors engineering should shift its focus from merely controlling human behavior to understanding and enhancing human resilience, especially regarding perception, decision-making, and adaptability. It is not enough to focus solely on prevention and control; we must also consider how individuals maintain resilience under conditions of family stress, unexpected events, and work fatigue. By incorporating AI into HRA from a resilience perspective, we can improve the overall resilience of the system. Furthermore, focusing solely on human error mechanisms is insufficient. A more reasonable model can only be achieved by integrating external systems, emphasizing the need for interdisciplinary development. The future lies in the convergence of AI, risk-informed decision-making, human performance models, and human reliability. In addition, safety research must draw insights from other fields to address complex challenges more effectively. Additionally, as highlighted by \cite{lye2024overview}, one area of potential future interest is the human reliability analysis for the Level 3 PSA during the decommissioning phase of the reactor.

\section{Declarations}

\subsection{Funding.} The research was supported by the National Natural Science Foundation of China (Grant No. T2192933).
\subsection{Author contribution.} Xiao Xingyu: Methodology, Software, Formal analysis, Data Curation, Visualization, Validation, Writing- Original draft preparation. Hongxu Zhu: Validation. Liang Jingang: Conceptualization, Resources, Supervision, Writing - Review and Editing, Project administration, Funding acquisition. Tong Jiejuan: Investigation, Supervision, Writing - Review and Editing. Wang Haitao: Supervision, Writing- Reviewing and Editing.

\bibliography{sn-bibliography}% common bib file

\end{document}